\documentclass[twocolumn, prl, superscriptaddress]{revtex4-1}
\usepackage{graphicx}
\usepackage{dcolumn}
\usepackage{bm}
\usepackage{hyperref}
\usepackage[colorinlistoftodos]{todonotes}
\usepackage{longtable}
\presetkeys{todonotes}{inline,backgroundcolor=yellow}{}

\newcommand{\ket}[1]{{\left| {#1} \right\rangle}}

\newcommand{\avg}[1]{\left\langle {#1}\right\rangle }

\newcommand{\mrm}[1]{\mathrm{{#1}}}

\newcommand{\Esca}{\mathcal{E}}

\newcommand{\ef}{\Esca_\mrm{eff}}

\begin{document}
\title{Ultracold mercury-alkali molecules for electron electric dipole moment searches}

\author{A. Sunaga}
\affiliation{Tokyo Metropolitan University, 1-1, Minami-Osawa, Hachioji-city, Tokyo 192-0397, Japan}

\author{V. S. Prasannaa}
\affiliation{Physical Research Laboratory, Atomic, Molecular and Optical Physics Division, Navrangpura, Ahmedabad-380009, India}


\author{M. Abe}
\affiliation{Tokyo Metropolitan University, 1-1, Minami-Osawa, Hachioji-city, Tokyo 192-0397, Japan}

\author{M. Hada}
\affiliation{Tokyo Metropolitan University, 1-1, Minami-Osawa, Hachioji-city, Tokyo 192-0397, Japan}

\author{B. P. Das}
\affiliation{Department of Physics, Tokyo Institute of Technology,
2-1-I-H86 Ookayama, Meguro-ku, Tokyo 152-8550, Japan}

\date{\today}

\begin{abstract}
Heavy polar diatomic molecules are the leading candidates in searches for the permanent electric dipole moment of the electron (eEDM). Next-generation eEDM search experiments ideally require extremely large coherence times, in large ensembles of trapped molecules that have a high sensitivity to the eEDM. We consider a family of molecules, mercury-alkali diatomics, that can be feasibly produced from ultracold atoms. We present calculations of the effective electric fields experienced by the electron in these molecules. The combination of reasonably large effective electric fields, and the possibility of obtaining trapped ultracold samples, leads us to identify these molecules as favorable candidates for eEDM search experiments.
\end{abstract}

\maketitle

The parity (P) and time reversal (T) violating electric dipole moment of the electron (eEDM) is one of the most important tabletop probes of physics beyond the standard model of elementary particles~\cite{DeMilleReview,ChuppReview,Fukuyama}. It can provide information on PeV scale physics, which is well beyond the reach of current accelerators~\cite{Nath}. Also, eEDM could offer insights into the baryon asymmetry in the universe~\cite{Fuyuto}. The leading candidates for eEDM searches are heavy polar diatomic molecules~\cite{Baron2014,Cairncross2017,Baron2017}. The current best upper bound on the eEDM is provided by ThO~\cite{Baron2014,Baron2017}, followed by limits from HfF$^+$~\cite{Cairncross2017} and YbF~\cite{Hudson2011}. A typical experiment measures energy shifts between different electron spin projections relative to the internuclear axis of a molecule -- using the  theoretically calculated value of the effective electric field ($\mathcal{E}_\mathrm{eff}$) experienced by the electrons in the molecule, the measured energy shifts can be related to the fundamental eEDM. There can also be energy shifts due to another P and T violating property, the scalar-pseudoscalar (S-PS) interaction between the electrons and the nuclei, parametrized by a theoretically calculated S-PS coefficient ($W_s$). The observation of a nonzero eEDM or S-PS energy shift could provide model-independent evidence of new physics beyond the standard model. 

Out of the plethora of polar molecules that are available for eEDM experiments, mercury-containing diatomics \cite{Kozlov2006,Prasannaa2015,Sasmal2016} are distinguished by their exceptionally large values of $\ef$ and $W_s$ compared to other analogous systems. For example, HgF \cite{Prasannaa2015} has a significantly larger $\ef$ even compared to molecules with mercury substituted by heavier atoms (e.g., RaF \cite{Kudashov2014}). The enhanced sensitivity of Hg-containing molecules derives from the contraction of the valence $6s_{1/2}$ and $6p_{1/2}$ orbitals due to the weaker screening by the outermost core $d$ electrons in Hg~\cite{Sunaga2018}. 

Beyond just the intrinsic sensitivity of a molecule to P and T violating physics determined by its $\ef$ and $W_s$ values, the sensitivity of an eEDM experiment improves with an increase in the electron spin coherence time and the total number of molecules observed during the experiment. Very long spin coherence times can be obtained with ultracold molecules trapped in optical dipole traps and optical lattices \cite{Park2017,Covey2018}. This strongly motivates the use of molecules whose electronic properties are amenable to direct laser-cooling(e.g.,\cite{Tarbutt2013,Isaev2010,Kozyryev2017,Lim2018}), or which can be assembled out of trapped ultracold atoms \cite{Meyer2009}. 

In this paper, 
we identify a set of Hg-containing molecules with high eEDM sensitivities 
: mercury-alkali diatomic molecules (HgA = HgLi, HgNa, and HgK). 
Experiments using these molecules have the potential to improve upon the current best eEDM measurements \cite{Cairncross2017,Baron2017} by at least one order of magnitude, with a commensurate increase in the energy scale up to which new physics effects can be probed. 

\emph{Theoretical calculations. --} The molecular properties of interest, $\mathcal{E}_\mathrm{eff}$ and $W_s$, are determined by the expressions~\cite{Abe,Barr} 
\begin{eqnarray}
\mathcal{E}_\mathrm{eff}&=& -2ic \sum_{j=1}^{N_e} \langle \psi \arrowvert \beta \gamma_5 \, p_j^2 \arrowvert \psi \rangle, \\
W_s&=&iG_F\sqrt{2} \sum_{j=1}^{N_e} \langle \psi \arrowvert \beta \gamma_5 \rho_A (\bm{r}_{Aj}) \arrowvert \psi \rangle,
\end{eqnarray}
where $\psi$ is the ground state wave function of a molecule, \textit{j} refers to summation over the electrons in the system, $\beta$ is the Dirac beta matrix, $\gamma_5$ is the product of the gamma matrices, \textit{p} the momentum operator for electron, $\rho_A$ the nuclear charge density, and $G_F$ is the Fermi coupling constant ($2.22249 \times 10^{-14}$ atomic units). We assume that only the ${}^{202}$Hg atom significantly affects $W_s$, as the contribution of the lighter atom is insignificant (cf.\ \cite{Sunaga2016}). 

We also calculate the molecular dipole moment (MDM), which is useful in determining the external electric field that one needs to apply, in order to polarize the molecule. The expression for the MDM of a molecule is

\begin{eqnarray}
\mathrm{MDM} = \langle \psi \arrowvert (-\sum_i r_i + \sum_A Z_{A} r_A) \arrowvert \psi\rangle   
\end{eqnarray}

In the above expression, the summation over the electronic coordinates
is given by $i$, while that over the nuclear coordinates is
indicated by $A$. Therefore, $r_i$ is the position vector from the origin
to the site of the $i^{th}$ electron, while $r_A$ is the position vector
from the origin to the coordinate of the $A^{th}$ nucleus. $r_A$, in our case, is the equilibrium bond length for the molecule HgA, with A $=$ Li, Na, or K, since we choose the Hg atom as our origin. $Z_A$ is the atomic number of the alkali atom, A. 

The properties given by equations (1), (2), and (3), can be obtained once we solve for the wave function, $\psi$. We employ a relativistic coupled cluster method, where the wave function is given by 

\begin{eqnarray}
\arrowvert \psi \rangle = e^T \arrowvert \Phi_0 \rangle
\end{eqnarray}

Here, $T$ refers to the cluster operator, and is associated with exciting holes (occupied orbitals) to particles (unoccupied ones). The exponential structure, $e^T$, takes into account all possible hole-particle excitations in the system, and $\ket{\Phi_0}$ is the Dirac-Fock (DF) reference determinant that it acts on. The DF method is the relativistic version of the Hartree-Fock approach, where each electron in a molecule experiences a mean potential due to the all the other electrons. The difference between the two-body Coulomb and the DF interactions is referred to as the residual interaction. The physical processes arising from the residual interaction are known as correlation effects. The coupled cluster method (CCM), which is considered to be the gold standard of many-body theory~\cite{LCT2}, is a powerful and efficient way of determining electron correlation. The CCM when compared to finite order many-body perturbation theory has the advantage of capturing the effects of the residual interaction to all orders in perturbation, for a given level of hole-particle excitation. It also fares better than the truncated configuration interaction (CI) method, another well-known approach that goes beyond the DF approximation where the wave function is written as a linear combination of several configuration states, in that for a given level of hole-particle excitation, the CCM includes more physical effects arising due to correlation~\cite{LCT2}. Also, unlike truncated CI, the coupled cluster is size extensive, that is, the energy scales with the number of particles. In such a framework, the most straightforward way to express an expectation value of an operator, $O$, is 

\begin{eqnarray}
\langle O \rangle = \frac{\langle \Phi_0 \arrowvert e^{T \dagger}Oe^T \arrowvert \Phi_0 \rangle}{\langle \Phi_0 \arrowvert e^{T \dagger}e^T \arrowvert \Phi_0 \rangle}
\end{eqnarray}

The above equation can be rewritten as follows~\cite{Bartlett} 

\begin{eqnarray}
\langle O \rangle &=& \langle \Phi_0 \arrowvert e^{T \dagger}O_Ne^T \arrowvert \Phi_0 \rangle_C \nonumber \\ &+& \langle \Phi_0 \arrowvert O \arrowvert \Phi_0 \rangle
\end{eqnarray}

The subscripts \textit{N} and \textit{C} refer to normal ordered arrangement of operators and connected terms respectively~\cite{Kvas,Bishop,Lindgren}. In our work, we consider single and double hole-particle excitations (the relativistic CCSD approximation~\cite{Eliav,Visscher}) in solving the coupled cluster equations, while for the expectation value, we only consider the terms that are linear in \textit{T} (the linear expectation value-CCSD or the LE-CCSD approximation). Therefore, the expression for the expectation value becomes

\begin{eqnarray}
\langle O \rangle &=& \langle \Phi_0 \arrowvert (1+T_1^{\dagger}+T_2^{\dagger})O_N(1+T_1+T_2) \arrowvert \Phi_0 \rangle_C \nonumber \\ &+& \langle \Phi_0 \arrowvert O \arrowvert \Phi_0 \rangle
\end{eqnarray} 

The validity of this approximation in calculating  $\mathcal{E}_\mathrm{eff}$ has been tested in a previous work~\cite{FFCC}. Although the previous work~\cite{FFCC} shows that the non-linear terms may contribute to MDM, the contribution of the higher order correlation would be small for the case of HgA, as shown later.

For DF computations and the atomic to molecular orbital transformations, we employed the UTChem code~\cite{Utchem,utchem2}, while the CCSD amplitudes were obtained from Dirac08~\cite{Dirac}. We then computed the CCSD expectation values using integrals and amplitudes from UTChem and Dirac08. 

For the DF calculation, optimized functions, called basis sets, are employed for each atom in a molecule. Among the simplest options is the Gaussian-type double zeta (DZ) basis~\cite{Dyall3}. The triple zeta (TZ) basis, an enlarged version of the DZ basis, is a better quality than the latter, followed by quadruple zeta (QZ) basis, and so on. More functions can be included in a basis, to take into account additional physical effects. We used uncontracted Dyall's triple zeta quality basis sets (more specifically the cvTZ basis~\cite{Dyall,Dyall2}, which includes additional polarizing functions) for all of the atoms in these molecules. In the CCSD calculations, we cut off the virtual spinors with orbital energy above 100 atomic units. We used the following bond lengths (in Angstroms): HgLi: 2.92, HgNa: 3.52, and HgK: 3.90~\cite{Thiel}. The direction of the MDM and the molecular axis are from the mercury atom to the alkali atom. 

\begin{table*}[t] 
\squeezetable
 \centering
 \begin{tabular}{cccccccccc}
 \hline
  Molecule & $\mathcal{E}_\mathrm{eff}^{DF}$&$W_s^{DF}$&$MDM^{DF}$&  $\mathcal{E}_\mathrm{eff}^{corr}$&$W_s^{corr}$&$MDM^{corr}$&$\mathcal{E}_\mathrm{eff}$&$W_s$&$MDM$ \\
 \hline
  HgLi&13.74&31.02&-1.47&24.05&55.35&1.95&37.79&86.37&0.48\\
  HgNa&7.59&17.15&-0.88&12.74&29.31&1.15&20.33&46.46&0.27\\
  HgK&5.73&12.95&-1.48&10.51&24.10&1.72&16.24&37.05&0.24\\
 \hline
 \end{tabular}
 \label{tab:contributions}
 \caption{The calculated values of $\mathcal{E}_\mathrm{eff}$ (in GV/cm), $W_s$ (in kHz) and the MDM (in Debye). The Dirac-Fock (DF, in superscript), the correlation (corr, in superscript), and the total (no superscript) contributions have been provided. The direction of the MDM is taken as the molecular axis from the mercury to the alkali atom.}
\end{table*}

The results of our calculations of $\mathcal{E}_\mathrm{eff}$, $W_s$, and the MDM are given in Table I. An interesting feature of these systems is the unusually large effect of electron correlations due to the van der Waals bonding in these molecules, which we have not observed in other eEDM candidates such as YbF or HgX (X=F, Cl, Br, and I)\cite{Abe,Prasannaa2015}. The electron correlations increase $\mathcal{E}_\mathrm{eff}$ and $W_s$ to almost thrice their DF values, while substantially affecting the MDM.  $\mathcal{E}_\mathrm{eff}$ and $W_s$ for all these molecules are comparable to these of YbF ($\mathcal{E}_\mathrm{eff}$ = 23.1 GV/cm~\cite{Abe}, $W_s$= -40.5 kHz~\cite{Sunaga2016}). 

We provide more detailed results in Table II-V, where we examine the individual correlation contributions from each term of Eq. (3). In Table II, we present the results for $\mathcal{E}_\mathrm{eff}$ and $W_s$. For brevity, we have used a notation where $OT_1$, for example, is actually $\langle \Phi_0 \arrowvert O_NT_1 \arrowvert \Phi_0 \rangle_C$, $T^\dagger_1OT_1$ is actually $\langle \Phi_0 \arrowvert T^\dagger_1O_NT_1 \arrowvert \Phi_0 \rangle_C$, and so on. The contribution from $OT_2$ (and its complex conjugate) is zero, due to the Slater-Condon rules. Also, $\langle \Phi_0 \arrowvert O_N \arrowvert \Phi_0 \rangle$ is zero, due to $O$ being in its normal-ordered form. 

\begin{table}[h!] 
\squeezetable
 \centering
 \begin{tabular}{ccccccc}
 \hline
  & &$\mathcal{E}_\mathrm{eff}$& & &$W_s$& \\
  Term &HgLi&HgNa&HgK&HgLi&HgNa&HgK\\
 \hline
  DF &13.74&7.59&5.73&31.02&17.15&12.95\\
  $OT_1$ + cc &21.46&11.42&9.24&49.28&26.22&21.15\\
  $T^\dagger_1OT_1$&1.10&0.62&0.56&2.45&1.39&1.25\\
  $T^\dagger_1OT_2$ + cc&2.96&1.46&1.28&7.03&3.48&3.02\\
  $T^\dagger_2OT_2$ &-1.47&-0.76&-0.56&-3.41&-1.78&-1.31\\
  \hline
 \end{tabular}
 \caption{Contributions from the individual terms of the LE-CCSD expression, to $\mathcal{E}_\mathrm{eff}$ (GV/cm) and $W_s$ (kHz) for HgLi, HgNa, and HgK; cc refers to complex conjugate of the term that it accompanies. The operator, $O$, can refer to either the operator of $\mathcal{E}_\mathrm{eff}$, or that of $W_s$, whose expectation value expressions are given in Eqs (1) and (2), respectively, in the main text. }
\end{table}

Table II shows that the correlation effects dominate in these systems, to an extent where the $OT_1$ term exceeds the DF value. This is in contrast to other eEDM candidates, such as YbF, BaF, or HgF, where correlation effects only slightly change the DF term (within 30 percent)~\cite{Abe,Sunaga2016,FFCC}. In HgF, for example, each of the terms involving $T$ do not exceed the DF value, and this combined with the fact that there are cancellations between the correlation terms, leaves behind a small correlation contribution (relative to the DF one)~\cite{VSP2}.

Tables III-V give the individual contributions to the MDM due to the electric term, which is the first term in Eq. (7). The third column is the sum of each electronic term; e.g. the third column for ``$T^\dagger_1H_{MDM}T_1$'' is the sum of electronic terms: ``DF'', ``$H_{MDM}T_1 $ + cc'', and ``$T^\dagger_1H_{MDM}T_1$''. 

We observe from Tables III-V that the contribution of the $H_{MDM}T_1 $ + cc  is much more dominant than the other correlation terms for all of the three molecules. The contributions from the $T_2$ terms are very small. From the maximum difference between the values including only $T_1$ (i. e. 0.47, 0.25, and 0.21 for HgLi, HgNa, and HgK, respectively) and the final values (i.e. 0.48, 0.27, and 0.24 for HgLi, HgNa, and HgK, respectively), we expect that the non-linear terms change the MDM by less than 0.03 D. Therefore the results at the LE-CCSD method are good estimates of the MDM of HgA molecules. Since the dominant correlation contribution to the MDM is from $H_{MDM}T_1$ + cc, we observe that low-order correlation effects are important in HgA. 

\begin{table}[h!] 
\squeezetable
 \centering
 \begin{tabular}{cccc}
 \hline
  Term &Electronic term&Sum&MDM\\
 \hline
  DF &-43.53&-43.53&-1.47\\
  $H_{MDM}T_1$ + cc &1.87&-41.66&0.40\\
  $T^\dagger_1H_{MDM}T_1$&0.06&-41.60&0.47\\
  $T^\dagger_1H_{MDM}T_2$ + cc&-0.04&-41.64&0.42\\
  $T^\dagger_2H_{MDM}T_2$ &0.05&-41.59&0.48\\
  \hline
 \end{tabular}
 \label{tab:contributions}
 \caption{Contributions from the electronic part of the CCSD linear expectation value, to the MDM of HgLi, in Debye (D). The third column refers to the sum of electronic terms. The term cc refers to the complex conjugate of the term that it accompanies. The nuclear contribution to the MDM is 42.06 D. }
\end{table}

\begin{table}[h!] 
\squeezetable
 \centering
 \begin{tabular}{cccc}
 \hline
  Term &Electronic term&Sum&MDM\\
 \hline
  DF &-186.80&-186.80&-0.88\\
  $H_{MDM}T_1$ + cc &1.11&-185.69&0.23\\
  $T^\dagger_1H_{MDM}T_1$&0.01&-185.68&0.25\\
  $T^\dagger_1H_{MDM}T_2$ + cc&-0.01&-185.69&0.23\\
  $T^\dagger_2H_{MDM}T_2$ &0.04&-185.66&0.27\\
 \hline
 \end{tabular}
 \label{tab:contributions}
 \caption{Contributions from the electronic part of the CCSD linear expectation value, to the MDM of HgNa, in Debye (D). The third column refers to the sum of electronic terms. The term cc refers to the complex conjugate of the term that it accompanies. The nuclear contribution to the MDM is 185.93 D. }
\end{table}

\begin{table}[h!]
\squeezetable
 \centering
 \begin{tabular}{cccc}
 \hline
  Term &Electronic term&Sum&MDM\\
 \hline
  DF &-357.30&-357.30&-1.48\\
  $H_{MDM}T_1$ + cc &1.68&-355.62&0.19\\
  $T^\dagger_1H_{MDM}T_1$&0.01&-355.60&0.21\\
  $T^\dagger_1H_{MDM}T_2$ + cc&-0.02&-355.63&0.19\\
  $T^\dagger_2H_{MDM}T_2$ &0.05&-355.57&0.24\\
 \hline
 \end{tabular}
 \label{tab:contributions}
 \caption{Contributions from the electronic part of the CCSD linear expectation value, to the MDM of HgK, in Debye (D). The third column refers to the sum of electronic terms. The term cc refers to the complex conjugate of the term that it accompanies. The nuclear contribution to the MDM is 355.81 D. 
}
\end{table}

We now estimate the errors in our calculations. The possible sources of the errors in our calculations of $\ef$ are due to three effects: (1) the non-inclusion of higher excitations in the wave function, for example, triples, (2) ignoring the non-linear terms in the coupled cluster operators in the expectation value, and (3) incompleteness of the basis functions. To estimate the error due to (1) and (2), we rely on comparisons between LE-CCSD and finite field CCSD(T) (FF-CCSD(T)) in a previous work.  In the finite field approach, a property is expressed as an energy derivative, rather than as an expectation value -- it therefore takes into account all the non-linear terms that are neglected in our expectation value approach. In our earlier work on HgF \cite{FFCC}, the largest change in $\ef$ between LE-CCSD and FF-CCSD(T) was approximately 5\%. We assume that the error due to (1) and (2) for HgA is comparable to HgF. For (3), we estimate that the error is $\sim$15\%, by examining the difference between results obtained with double zeta (cvDZ) and triple zeta (cvTZ) basis sets~\cite{Asunaga3}. This estimate assumes that the change from triple zeta (cvTZ) to quadruple zeta (cvQZ) quality basis sets is not larger than 15\% -- the assumption was tested for HgA using DZ, TZ, and QZ basis functions (but without the polarizing functions) where we observed that the difference between TZ and QZ basis sets was smaller than that between DZ and TZ basis sets~\cite{Asunaga3}. We combine these systematic error estimates linearly, and conservatively estimate a total error of 20\% in our calculations of $\ef$. Based on similar considerations, we do not expect the error in $W_s$ to be greater than 20\% either. 
From the expansion of the expectation value in Table III-V, we see that the error in the MDM due to the exclusion of higher-order excitations as follows: 1) the non-linear terms may not contribute to more than 0.03 D, and 2) the higher excitations like triples will not alter the MDM noticeably. Calculations of the MDM can be quite sensitive to the choice of basis, especially for molecules with van der Waals bonds such as HgA. Our results for the MDM of HgA molecules are in broad agreement with Cremer \emph{et al.} \cite{Cremer}, who used similar equilibrium bond lengths in their calculations but different basis sets and computational methods. Their basis sets were Dyall's DZ for the Hg core, Dyall's TZ for the valence Hg orbitals, aug-ccpVTZ for Li and Na, and 6-311 + + G(3df) basis for K; in addition, their calculations only considered scalar relativistic effects. Here, in comparison, we use Dyall's cvTZ basis sets and the Dirac-Coulomb Hamiltonian throughout.

\emph{Experimental aspects. --} In this section, we briefly comment on the possibility of an eEDM experiment, using HgA systems, based on some preliminary considerations. The figure of merit for the statistical sensitivity of an eEDM experiment using molecules is $F = \ef \, \sqrt{N \tau}$, where $N$ is the number of molecules interrogated in the experiment and $\tau$ is the coherence time for the electron spin precession. The values of $\ef$ for HgA molecules are comparable to, or larger than, those of some other molecules planned for use in next-generation eEDM experiments (cf.\ \cite{Lim2018,Aggarwal2018,Vutha2018}). In a possible optical lattice eEDM experiment with HgA,  a fairly large coherence time can be expected \cite{Park2017}. Using the computed MDM values, we estimated the magnitude of the laboratory electric field required to significantly polarize HgA molecules $\Esca_\mrm{pol} = 2 B_e/D$ (where $B_e$ is the equilibrium rotational constant of the molecule and $D$ is the MDM). The values of $\Esca_\mrm{pol}$ are (71,28,17) kV/cm for (HgLi, HgNa, HgK), implying that one can feasibly polarize a sample of trapped ultracold HgA molecules. It may be possible to apply large external fields without any significant leakage-current-induced spurious magnetic fields \cite{Ragba2018}. Cooling alkali atoms to micro-kelvin temperatures and trapping in optical lattices have been implemented for Hg atoms in the context of optical lattice clocks \cite{Katori,Yi2011}. Methods for assembling molecules from ultracold atoms have advanced significantly over the last decade \cite{Takekoshi2014,Park2015,Moses2015,Guo2016,Rvachov2018}. In particular, molecules isoelectronic to HgA have been produced at ultracold temperatures \cite{Roy2016,Barbe2018}, and methods for producing other ultracold alkali-alkaline earth molecules \cite{Devolder2018} (including a Hg-alkali molecule, HgRb \cite{Witkowski2017,Borkowski2017}) are currently being investigated. It seems within the realm of possibility that these techniques can be extended to the analog molecules HgLi, HgNa and HgK. We base our estimate for the eEDM sensitivty on the conservative assumption that $N = 10^4$ ultracold HgA molecules can be produced in an optical lattice, using the numbers demonstrated with isoelectronic YbLi molecules \cite{Hara2011,Hansen2013}. Based on the very large coherence times between hyperfine states that have been observed with lattice-trapped ultracold polar molecules \cite{Park2017}, we assume that a coherence time $\tau$ = 1 s can be realized. From numerical calculations of the Stark effect in the hyperfine and rotational states in HgA, we estimate that an electron spin orientation factor, $\xi = \avg{\vec{S}\cdot \hat{n}}$ = 0.13~\cite{ACVpriv}, can be achieved using lab electric fields of magnitude $\Esca_\mrm{lab} = \Esca_\mrm{pol}$. With these values of $N,\tau,\xi$ and $\ef$ and a total integration time of $T = 10^7$ s, we estimate preliminary eEDM sensitivities $\delta d_e = (1.3, 2.5, 3.1) \times 10^{-30} \ e$ cm for (HgLi, HgNa, HgK), offering the prospect of improvements over the current experimental limit ($|d_e| < 1.1 \times 10^{-29}$ e cm~\cite{Baron2017}).

To conclude, we have presented the results of our CCSD calculations of $\mathcal{E}_\mathrm{eff}$, $W_s$, and MDMs of Hg-alkali systems. Also, we present preliminary estimates of the expected  sensitivities for Hg-alkali molecules, which suggest that these systems could be promising candidates for eEDM experiments. Further work on the experimental aspects would be necessary in the future to explore the possibilities of performing eEDM experiments using HgA systems.

\emph{Acknowledgments. --} The authors would like to acknowledge Prof. A. C. Vutha for his useful suggestions and motivating the experimental aspects of this work. Computations were performed on the machine at Tokyo Metropolitan University, and on the GPC supercomputer at the SciNet Consortium~\cite{Loken2010}. This research was supported by JST, CREST, MEXT, NSERC and Canada Research Chairs. This work was supported by JSPS KAKENHI Grant No. 17J02767, No. 17H03011, No. 17H02881, and No. 18K05040.

\end{document}